\documentclass[runningheads]{llncs}
\usepackage[T1]{fontenc}
\usepackage{graphicx,verbatim}
\usepackage{amsmath,amsfonts}
\usepackage[most]{tcolorbox}
\usepackage{bm}
\usepackage{enumitem}
\usepackage{xcolor}
\definecolor{mypink}{RGB}{255,105,180} 
\usepackage[colorlinks=true,urlcolor=mypink]{hyperref}

\begin{document}
\title{Extracting Root-Causal Brain Activity Driving Psychopathology from Resting State fMRI}
\titlerunning{Root-Causal Brain Activity}
%

\author{Eric V. Strobl}  
\authorrunning{Strobl}
\institute{University of Pittsburgh}
  
\maketitle              
\begin{abstract}
Neuroimaging studies of psychiatric disorders often correlate imaging patterns with diagnostic labels or composite symptom scores, yielding diffuse associations that obscure underlying mechanisms. We instead seek to identify \emph{root-causal maps}---localized BOLD disturbances that initiate pathological cascades---and to link them selectively to symptom dimensions. We introduce a bilevel structural causal model that connects between-subject symptom structure to within-subject resting-state fMRI via independent latent sources with localized direct effects. Based on this model, we develop SOURCE (Symptom-Oriented Uncovering of Root-Causal Elements), a procedure that links interpretable symptom axes to a parsimonious set of localized drivers. Experiments show that SOURCE recovers localized maps consistent with root-causal BOLD drivers and increases interpretability and anatomical specificity relative to existing comparators.
\end{abstract}

%
%
\section{Introduction}
Complex psychiatric disorders such as schizophrenia and major depression alter brain activity measurable with functional magnetic resonance imaging (fMRI). Clinicians often compress heterogeneous symptom domains into diagnostic labels or total scores \cite{Newson20}, and neuroimaging studies commonly correlate regional or network connectivity with these composites \cite{Kaiser15,Li17}. However, because these aggregates do not align cleanly with localized dysfunction, analyses often yield diffuse connectivity associations that capture global complexity but do little to pinpoint isolated mechanistic components of psychopathology \cite{Feczko19,Segal23}.

Here, we instead aim to identify \emph{root-causal maps}, which we define as localized blood-oxygen-level-dependent (BOLD) disturbances that are \emph{directly} driven by independent sources and that subsequently propagate to produce downstream, network-wide abnormalities. In principle, Linear Non-Gaussian Acyclic Model (LiNGAM)--based causal discovery can recover the independent sources by exploiting independent component analysis (ICA) \cite{Shimizu06,Xu14,Strobl22,Strobl24_counter}. In voxelwise fMRI, however, standard ICA pipelines face practical barriers to both causal and anatomical interpretability: analyses often operate in an undercomplete regime, and the estimated mixing maps tend to be spatially distributed rather than focal \cite{Damoiseaux06}. More broadly, component-based decompositions such as spatial or temporal ICA, nonnegative matrix factorization (NMF) \cite{Sotiras15}, and dictionary learning \cite{Eavani12}, including their modern variants (e.g. \cite{Du20,Ha23,Li16}), do not separate direct source-to-voxel effects from indirect propagation through voxel-to-voxel interactions. As a result, they are not designed to localize upstream voxels in a causal sense.

We make three \textbf{contributions} toward isolating root-causal maps:
\begin{enumerate}[leftmargin=*]
\item We introduce a bilevel structural causal model that links localized root-causal maps to specific symptom outcomes in resting-state fMRI (rs-fMRI).
\item We develop SOURCE (Symptom-Oriented Uncovering of Root-Causal Elements), which identifies root-causal maps from diverse rs-fMRI datasets by jointly learning: (a) independent components that represent exogenous latent sources; (b) spatially localized \emph{root-proximal} maps for each source, enriched for direct source-to-voxel effects on brain activity; and (c) symptom axes that link selectively and parsimoniously to a small subset of sources.
\item We show that SOURCE recovers localized maps consistent with exogenous, independent drivers in multiple real datasets, outperforming existing methods in interpretability and localization.
\end{enumerate}
SOURCE thus improves our ability to isolate the neural origins of complex psychiatric phenotypes by enabling root-causal analysis of rs-fMRI.

\section{Bilevel Causal Model}
\textit{Structural causal models} (SCMs) represent $m$ observed variables $\bm Z$ via structural assignments
\[
Z_i=f_i(\mathrm{Pa}(Z_i),E_i),\qquad i=1,\dots,m,
\]
where $\mathrm{Pa}(Z_i)\subseteq \bm Z\setminus\{Z_i\}$ denotes the set of direct causes (parents) of $Z_i$ and $E_i$ captures exogenous variation. The induced directed graph $\mathbb G$ includes an edge $Z_j\to Z_i$ whenever $Z_j\in \mathrm{Pa}(Z_i)$; unlike standard acyclic SCMs, we permit $\mathbb G$ to contain directed cycles.

We partition $\bm Z$ into two non-overlapping blocks: voxel-level summaries $\bm X\in\mathbb R^{p}$ and symptom measures $\bm Y\in\mathbb R^{q}$. We also introduce subject-level driver variables $\bm S\in\mathbb R^{K}$. We then specify a two-level model that links \emph{between-subject} variation to \emph{within-subject} dynamics. At the between-subject level, latent drivers $\bm S$ affect voxels $\bm X$ both directly and indirectly through voxel-to-voxel propagation, and a subset of voxels subsequently shapes the symptom structure $\bm Y$. At the within-subject level, mutually independent latent sources drive voxel time series through localized source-to-voxel effects and propagation over a fixed voxel interaction graph. We further specify the model below.

\subsection{Between Subjects}
We residualize $\bm X$ and $\bm Y$ for standard covariates (e.g., site/scanner, age, sex, motion summaries, medication class/dose) and keep the same symbols for the residualized quantities. We treat $\bm S$ as exogenous and posit the SCM
\begin{equation}\nonumber
\begin{aligned}
\bm X &= \bm S\Gamma+\bm X B+\bm E_X, 
&& \Gamma\in\mathbb R^{K\times p},\ B\in\mathbb R^{p\times p},\\
\bm Y &= \bm X\Phi+\bm E_Y, 
&& \Phi\in\mathbb R^{p\times q},
\end{aligned}
\end{equation}
where $B$ encodes (possibly cyclic) voxel-to-voxel propagation with spectral radius less than one. The SCM enforces biologically plausible directions: latent sources act on voxels via $\Gamma$, voxels influence voxels via $B$, and voxels provide the direct inputs to symptoms via $\Phi$. Any symptom axis $y=\bm Y\alpha$ satisfies $y=\bm X(\Phi\alpha)+e_\alpha$.

We call a voxel \emph{root-causal} for a symptom axis $y=\bm Y\alpha$ if it is (i) directly impacted by at least one driver in $\bm{S}$, and (ii) lies on a directed path to $y$ in the augmented graph that includes voxel-to-symptom edges weighted by $\Phi\alpha$. This targets neural causes as far upstream as rs-fMRI voxels allow.

\subsection{From Between to Within Subjects}
For subject $i$, each between-subject driver $S^{(i)}_j$ summarizes a within-subject latent time series $F^{(i)}_j(\cdot)$:
\[
S^{(i)}_j=\mathcal G_\theta\!\big[\,F^{(i)}_j(t);\ t\in\mathcal T_i\big],
\]
where $\mathcal G_\theta$ aggregates a time course to a scalar (e.g., ALFF$_{0.01\text{--}0.10}$, fALFF, $\log\!\operatorname{Var}$), and $\theta$ collects preprocessing/band choices. Thus $\bm S$ lives between subjects, while $\bm F(t)$ lives within subjects.

\subsection{Within Subjects: Localized SCM}
Standard LiNGAM assumes each observed variable has its own statistically independent error term \cite{Shimizu06}. Voxelwise, this would require an independent exogenous disturbance for every voxel, which is hard to reconcile with fMRI’s strong spatial dependence. We instead posit a low-dimensional set of exogenous drivers: mutually independent latent sources $\bm F(t)\in\mathbb R^{1\times K}$ with
$p(F_1,\dots,F_K)=\prod_{k=1}^K p(F_k)$.
Each source $F_i(t)$ acts \emph{directly} on voxels through a localized spatial map $\Gamma_{i \cdot}$, so spatial dependence is captured by the structure of $\Gamma$.

For subject $i$ at time $t$, let $\bm X^{(i)}(t)\in\mathbb R^{1\times p}$ be voxel signals and $\bm F^{(i)}(t)$ the corresponding source activations. Let $B$ encode instantaneous voxel-to-voxel propagation. We posit
\begin{equation}\nonumber
\bm X^{(i)}(t)=\bm X^{(i)}(t)B+\bm F^{(i)}(t)\Gamma^{(i)}+\bm\delta^{(i)}+\bm\varepsilon^{(i)}(t),
\end{equation}
where $\Gamma^{(i)}\in\mathbb R^{K\times p}$ are direct source-to-voxel effects, $\bm\delta^{(i)}$ is a subject intercept, and $\bm\varepsilon^{(i)}(t)$ is noise. Rearranging yields a mixing form with template
$\bm M^{(i)}=\Gamma^{(i)}(I-B)^{-1}$. After centering and voxelwise standardization, we assume cross-subject heterogeneity reduces to column scalings, so all subjects share a common template $\bm M$:
\[
\bm X^{(i)}(t)=\big(\bm F^{(i)}(t)-\bm m^{(i)}\big)\bm M+\widetilde{\bm\varepsilon}^{(i)}(t),\qquad \bm M=\Gamma(I-B)^{-1},
\]
where $\bm m^{(i)}=\mathbb E_t[\bm F^{(i)}(t)]$ denotes subject $i$'s mean source activation across time, so $\bm F^{(i)}(t)-\bm m^{(i)}$ is the centered source time course. Pooling standardized time points across subjects then supports ICA-based recovery of source time courses (up to scale/permutation).

\section{Algorithm}
We aim to (i) estimate spatially localized \emph{root-proximal} maps that target direct source-to-voxel effects in $\Gamma$ from within-subject rs-fMRI while treating the high dimensional $B$ matrix as a nuisance, and (ii) identify which between-subject summaries $\bm S^{(i)}=\mathcal G_\theta[\bm F^{(i)}(\cdot)]$ directly cause symptom axes $y=\bm Y\alpha$. SOURCE follows a three-stage pipeline: (1) recover latent sources, (2) select a compact set of root-proximal voxels that best anchor those sources, and (3) learn a symptom axis as a sparse function of a small number of sources.

\subsection{Recovering Latent Sources}
We first regress within-subject nuisance terms (including CSF mean and drift trends) from each subject’s GM-masked voxel time series, then z-score each voxel’s time course within subject. Next, we concatenate time points across subjects, apply spatial smoothing (e.g., 6 mm FWHM) to the concatenated data in subject-wise blocks, and run ICA. Finally, we filter components to remove noise-like patterns and retain candidate sources $\bm{F}$.

\subsection{Identifying Root-Proximal Voxels}
Let $\widetilde X_i(t)$ denote a spatially smoothed voxel series. We avoid conditioning on all other voxels in regression (computationally and statistically unstable) and instead approximate local confounding by conditioning on a \emph{center-excluded Gaussian neighborhood}. With kernel center weight $w_0$,
\[
N_i(t)=\frac{\widetilde X_i(t)-X_i(t)w_0}{1-w_0},
\]
which removes self-leakage. For each voxel $i$, we residualize (a) ICA source time courses and (b) $\widetilde X_i(t)$ on $(1,N_i(t))$, yielding $\bm F^{-i}(t)$ and $\widetilde X_i^{-i}(t)$. We then compute
\[
\theta_{\cdot i}=\mathrm{cor}\!\big(\bm F^{-i}(t),\widetilde X_i^{-i}(t)\big),\qquad
\eta_{\cdot i}=\mathrm{cor}\!\big(\bm F(t),\widetilde X_i(t)\big).
\]

Because the product $\eta_{j\cdot}\odot\theta_{j\cdot}$ is dense and noisy, we estimate a sparse, spatially coherent effect map $\zeta_j\in\mathbb R^p$ for each source $j$ by fitting $\zeta_j\approx \eta_{j\cdot}\odot\theta_{j\cdot}$ with graph smoothness and sparsity:
\[
\widehat\zeta_j=\arg\min_{\zeta_j}\ \|\eta_{j\cdot}\odot\theta_{j\cdot}-\zeta_j\|_2^2
+\lambda_1\,\mathrm{TV}_G(\zeta_j)+\lambda_2\|\zeta_j\|_1,
\]
where $\mathrm{TV}_G$ is total variation on a voxel adjacency graph. We perform iteratively re-weighted least-squares (IRLS) quadratic majorization \cite{Daubechies10} using reweighted Laplacian and diagonal penalties, which preserves sparsity while encouraging piecewise-smooth spatial structure. We interpret $|\widehat\zeta_j|$ as a \emph{root-proximity} map: it highlights voxels whose source association is present marginally ($\eta_{ji}$) and persists after local adjustment ($\theta_{ji}$), downweighting effects attributable to propagation without explicitly estimating $B$. Note that $\eta_{ji}\theta_{ji} \not = 0$ is not a certificate that $\Gamma_{ji}\neq 0$, but a tractable surrogate that is empirically enriched for direct effects.

We tune $(\lambda_1,\lambda_2)$ to favor maps that are simultaneously smooth and spatially compact by maximizing
$\rho=\frac{1}{K}\sum_{j=1}^K (1-R_j)(1-D_j)$,
where
\[
R_j=\frac{\widehat\zeta_j^\top L\,\widehat\zeta_j}{\lambda_{\max}(L)\,\widehat\zeta_j^\top \widehat\zeta_j}\in[0,1]
\]
is the Rayleigh--Ritz normalized Laplacian roughness score with normalized graph Laplacian $L$, and
$D_j=\|\widehat\zeta_j\|_1^2/(p\|\widehat\zeta_j\|_2^2)$ measures diffuseness. The product $(1-R_j)(1-D_j)$ rewards solutions only when both criteria hold.

\subsection{Aligning Multiple Symptom Axes}
We learn a symptom axis $y=\bm Y\alpha$ that aligns with a sparse subset of subject-level drivers (source-derived summaries) by solving
\[
\arg\max_{\alpha\ge 0,\ \beta}\ \mathrm{cor}(\bm S\beta,\bm Y\alpha)-\lambda_3\|\beta\|_0
\quad\text{s.t.}\quad
\mathrm{Var}(\bm S\beta)=1,\ \|\alpha\|_1=1,
\]
which fixes scale and enforces non-negativity over $\alpha$ so that $\bm Y \alpha$ remains an easily interpretable severity score. To extract multiple nonredundant axes, one can iteratively deflate $\bm S$ by regressing out previously found projections $\bm S\widehat\beta$ and re-solving. In the population linear no-confounding regime with $\bm S$ upstream of $\bm Y$, correlation-based extraction operates entirely within the $\bm Y$-predictive, and hence ancestor, subspace of $\mathrm{span}(\bm S)$; deflation then sequentially removes extracted $\bm Y$-predictive directions to reveal additional ones.

For each symptom projection, SOURCE returns (i) a sparse set of independent sources associated with and ancestral to that symptom dimension and (ii) corresponding spatially localized root-proximity maps that help localize the neural drivers of symptom expression. Overall, \textsc{SOURCE} is dominated by the IRLS routine and scales as $\mathcal{O}\!\left(G\,K\,T_{\mathrm{IRLS}}\,p^3\right)$, where $G = |\Lambda_{1}|\cdot|\Lambda_{2}|$ is the number of $(\lambda_{1},\lambda_{2})$ grid pairs and $T_{\mathrm{IRLS}}$ is the number of reweighting iterations. The $p^3$ term is a conservative worst-case upper bound for a sparse $p\times p$ Cholesky factorization \cite{Chen08}; in practice, the Laplacian-based sparsity structure typically leads to substantially lower runtimes.
Code link: \url{github.com/ericstrobl/SOURCE}.

\section{Experiments}\subsection{Setup}
\subsubsection{Comparators} We compared SOURCE against three representative baselines spanning modern ICA, non-negative matrix factorization (NMF), and sparse dictionary learning. Unless the authors recommended a certain dimensionality, we extracted 100 components to capture mesoscale functional networks:
\begin{enumerate}[leftmargin=*]
    \item \textbf{Spatially Constrained ICA (scICA)} \cite{Hesse06,Du20}: temporal ICA under a linear SCM and imposing a soft spatial prior on the mixing matrix using the NeuroMark v1.0 template, fixing the model order to $K=53$. We pooled subjects by stacking time points to increase the effective sample size and improve causal discovery \cite{Xu14,Sanchez19}.
    \item \textbf{Orthogonal Projective NMF (opNMF)} \cite{Ha23}: an NMF variant that enforces orthonormal components and projective coefficients to promote sparse, localized networks.
    \item \textbf{Rank-One Dictionary Learning (r1DL)} \cite{Li16}: alternates least-squares updates of a unit-norm temporal vector and sparse spatial loading with deflation to extract multiple networks; we use the recommended $r=0.07$ sparsity and 80 components.
\end{enumerate}
scICA, opNMF, and r1DL do not learn outcome measures, so we paired each method with a total severity score and, when applicable, every available subscore. We additionally evaluated \textbf{three ablations of SOURCE}: (i) \emph{no root-proximal map}, using only unconditional correlation; (ii) \emph{no correlation maximization}, correlating only with the total score; and (iii) both (i) and (ii). We evaluated all methods using 100 bootstrap resamples per dataset on the metrics defined below.

\subsubsection{Metrics} We first evaluated each algorithm's predictive performance using the standard Pearson \textbf{correlation} coefficient on the held-out bootstrap test sets. SOURCE and SOURCE without root-proximal maps attain the same correlation because this metric captures predictive accuracy but is insensitive to spatial localization. To quantify localization---and because our primary goal is to isolate a small set of root-proximal voxels---we also evaluated how well each method preserves symptom--brain predictivity while concentrating effects spatially. Our primary metric is \textbf{correlation density},
\begin{equation} \nonumber
\mathrm{CD}
=
\widehat{\mathrm{cor}}\!\left(\bm S\widehat{\bm\beta},\,\widetilde{\bm Y}\right)
\Big/
\left(
\frac{1}{p}\sum_{i=1}^{p}\sum_{k=1}^{K}
\left|\widehat \chi_{k i}\right|\,\left|\widehat\beta_{k}\right|
\right),
\end{equation}
where $\widetilde{\bm{Y}}$ is the (fixed or learned) target and $\widehat{\chi}\in\mathbb R^{K\times p}$ is the method-specific effect map, with each row normalized to unit $\ell_2$ norm for comparability across methods. CD reports held-out symptom--brain correlation per unit average symptom-weighted effect magnitude.

We also report \textbf{compactness} of the symptom-weighted effects,
\begin{equation} \nonumber
  \mathrm{Compactness}
  =
    \frac{\max_{v \in \mathcal{V}} \sum_{u \in \mathcal{N}_6(v)} \bigl| \widehat{\chi}_{\cdot u}^{\top} \widehat{\beta} \bigr|}{
    \sum_{u \in \mathcal{V}} \bigl| \widehat{\chi}_{\cdot u}^{\top} \widehat{\beta} \bigr|},
  \label{eq:compactness}
\end{equation}
where $\mathcal{N}_6(v)$ is the 6-connected neighborhood of $v$, including $v$ itself. Compactness equals $1$ for a single hotspot and decreases toward $0$ as effects spread. Finally, we recorded the \textbf{run time} of SOURCE and evaluated the \textbf{neuroanatomical plausibility} of its outputs.

\subsubsection{Preprocessing and hyperparameters} We applied a standard FSL-based preprocessing pipeline to rs-fMRI \cite{Jenkinson12}, producing preprocessed BOLD images in MNI152 space at 3\,mm resolution with subject-specific tissue masks. We built group gray matter (GM) and cerebrospinal fluid (CSF) masks by retaining voxels present in $\ge 80\%$ of subjects, extracted GM-masked voxel time series, and computed each subject's mean CSF signal (and first difference) as additional nuisance regressors. We performed within-subject nuisance regression (motion, aCompCor, CSF terms) with linear/quadratic detrending. For model estimation, we maximized $\rho$ over $\lambda_1,\lambda_2 \in \{1, 0.1, 0.01, 0.001\}$ and selected $\lambda_3$ by cross-validation over $\{1, 1/2, 1/4, 1/8, 1/16\}$ to maximize correlation for each score.

\subsection{Psychosis}

\begin{figure*}[!t]
    \centering
    \includegraphics[width=1\linewidth]{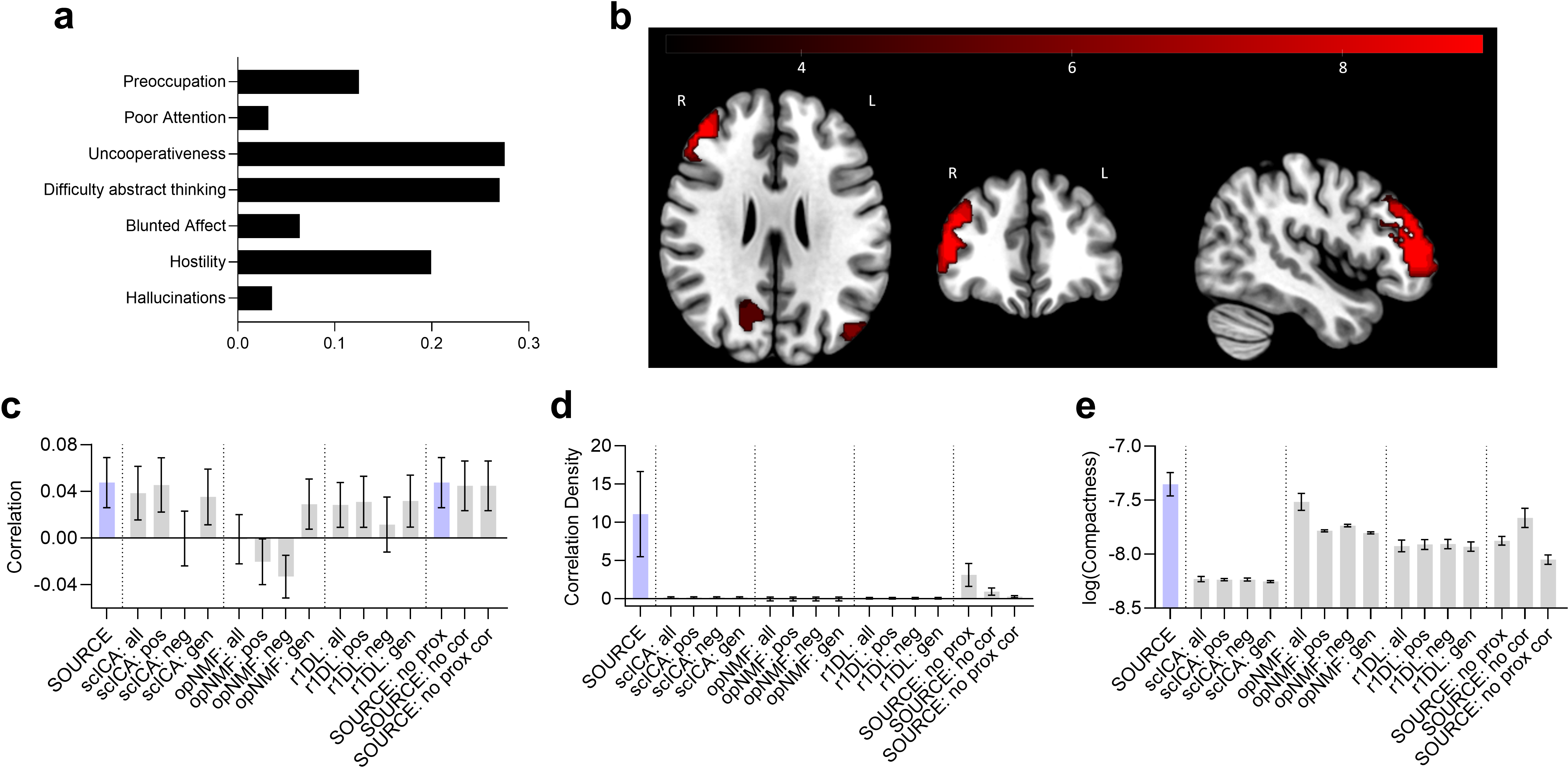}
    \caption{\textbf{Psychosis results.} (a) Learned PANSS axis (non-zero loadings). (b) rDLPFC localization. (c) Correlation. (d) Correlation density. (e) Compactness. Error bars denote 95\% confidence intervals of the mean.}
    \label{fig:BSNIP2}
\end{figure*}

We analyzed BSNIP2 (NIMH Data Archive; NDA ID: 2165) \cite{Clementz22}, restricting to 216 subjects with baseline rs-fMRI and complete item-level PANSS ratings (30 items as outcomes). SOURCE learned a symptom axis consistent with oppositional rigidity and preoccupation (7 non-zero item loadings; Figure \ref{fig:BSNIP2}a). The root-proximity maps localized the associated sources predominantly to the right dorsolateral prefrontal cortex (rDLPFC; Figure \ref{fig:BSNIP2}b), consistent with its role in rule maintenance and response switching \cite{Friedman22}. SOURCE matched the best held-out correlation while substantially improving correlation density and compactness (Figure \ref{fig:BSNIP2}c--e). We evaluated performance with the PANSS total score (all) and its three subscales: positive (pos), negative (neg), and general psychopathology (gen). SOURCE thus achieved the highest accuracy and parsimony. The algorithm also completed in $\sim$4 hours.

\subsection{Depression}

\begin{figure*}[!t]
    \centering
    \includegraphics[width=1\linewidth]{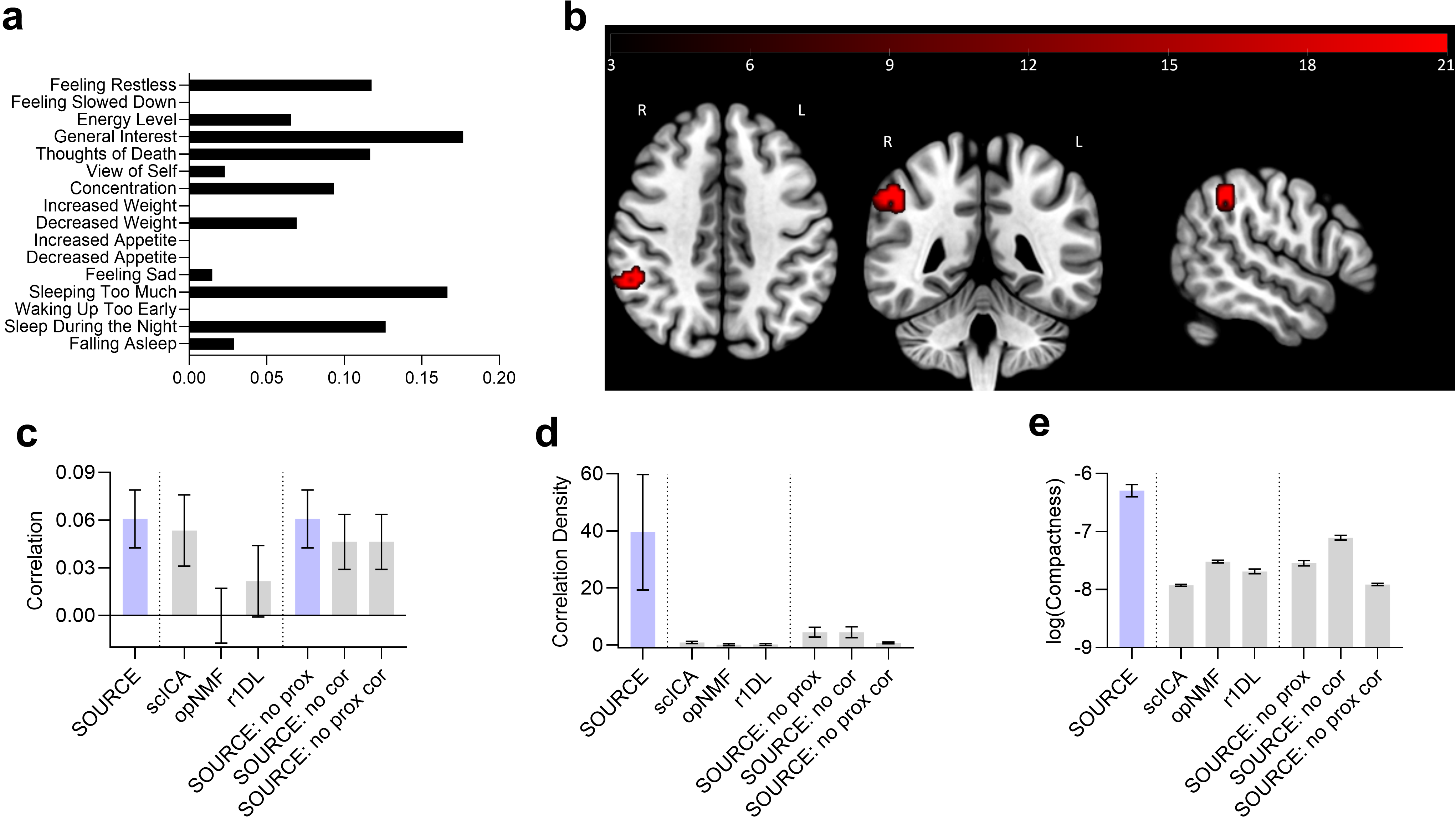}
    \caption{\textbf{Depression results.} (a) Learned symptom axis. (b) Localization. (c--e) Correlation, correlation density and compactness.}
    \label{fig:EMBARC}
\end{figure*}

We analyzed EMBARC (NDA ID: 2199) \cite{Trivedi16}, retaining 281 participants with QC-passed rs-fMRI and complete baseline QIDS-SR items. SOURCE recovered an anergia/neurovegetative symptom axis (Figure \ref{fig:EMBARC}a) explained by a single source, localized to the right inferior parietal lobule (Figure \ref{fig:EMBARC}b), consistent with frontoparietal control and attentional-effort circuitry implicated in depression \cite{Schultz18,Kaiser15,Alahmadi21,Husain07}. SOURCE again matched the highest correlation, and substantially outperformed all methods on correlation density and compactness (Figure \ref{fig:EMBARC}c--e); ablations indicated that identifying root-proximal voxels and learning the symptom axis both increased performance. Runtime was $\sim$2h15m.

\section{Conclusion}

We introduced a bilevel linear non-Gaussian SCM for rs-fMRI that links within-subject latent spatial sources to between-subject symptom structure while treating voxel-to-voxel propagation as an invariant nuisance. SOURCE operationalizes this model by recovering candidate sources, estimating sparse root-proximal maps, and learning interpretable item-level symptom axes driven by a small subset of sources. Across datasets, SOURCE produces anatomically localized signatures with strong predictive performance and improved interpretability relative to component-based baselines.



%
%
%
\bibliographystyle{splncs04}
\bibliography{biblio}

\end{document}